\documentclass{ws-p8-50x6-00}

\begin{document}

\title{Observation of eta-mesic nuclei in photoreactions:
results and perspectives}

\author{G.A. Sokol, A.I. L'vov and L.N. Pavlyuchenko}

\address{
   Lebedev Physical Institute, Leninsky Prospect 53, Moscow 117924, Russia \\
   E-mail: gsokol@x4u.lebedev.ru}

\maketitle

\abstracts{
Recent results from the LPI experiment on searching for $\eta$-mesic
nuclei in photoreactions are discussed and further perspectives are
summarized.}

This talk concerns experimental investigations of new objects of the
nuclear physics, $\eta$-mesic nuclei ($_\eta A$) which are bound
systems of the $\eta$-meson and a nucleus.  Discovery and learning
properties of these objects are of fundamental significance for
understanding interactions of the $\eta$-meson with nucleons and
nucleon resonances, for understanding a behavior of hadrons in nuclear
matter.

\begin{figure}[hbt]
\centerline{\epsfxsize=0.7\textwidth \epsfbox{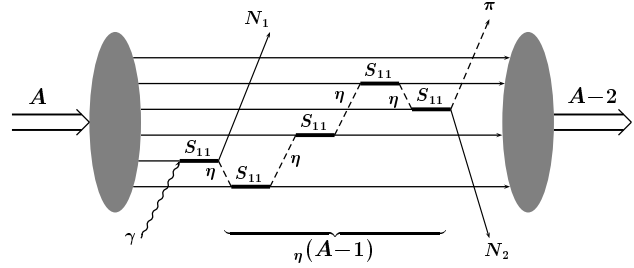}}
\caption{Mechanism of creation and decay of an ${\eta}$-mesic nucleus
and its decay into a $\pi N$ pair.}
\label{fig:mechanism}
\end{figure}

A mechanism of creation of $\eta$-mesic nuclei in photoreactions and
their decay into a $\pi N$ pair is shown in Fig.~\ref{fig:mechanism}.
At the first stage of the reaction, an incoming photon produces a slow
$\eta$-meson and a fast nucleon $N_1$ escaping the nuclear target.
Then the $\eta$-meson propagates in the nuclear matter and undergoes
multiple rescattering off nucleons.  At last, $\eta$ annihilates
producing an $\pi N_2$ pair escaping the nucleus.  The $S_{11}(1535)$
nucleon resonance plays a fundamental role in all that dynamics. It
mediates creation and annihilation of $\eta$ and it makes the
$\eta$-meson bound in the nucleus due to an $S_{11}(1535)$-induced
$\eta N$ interaction (attraction).  Knowing energies/momenta of the
ingoing ($\gamma$) and outgoing ($N_1$, $\pi N_2$) particles, one can
determine energies/momenta of the $\eta$ and $S_{11}(1535)$ propagating
in the nuclear matter.  The width and binding energy of $\eta$ in the
nucleus can be determined as well.

\begin{figure}[ht]
\psfig{file=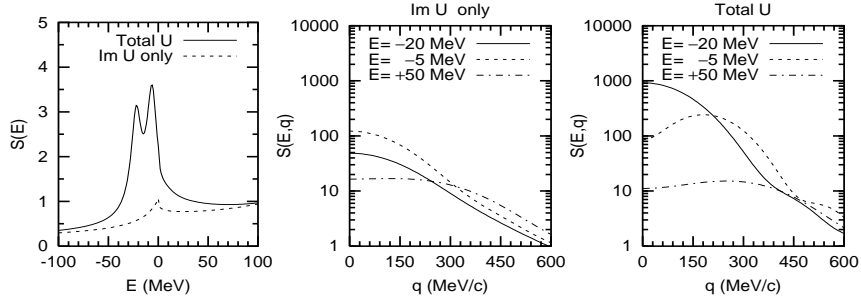,width=\textwidth,height=4.0cm}
\caption{Spectral functions $S(E)$ and $S(E,q)$ (in arbitrary units)
of the (kinetic) energy $E$ and momentum $q$ of $\eta$ in the nucleus.
They are found with a rectangular-well optical potential
simulating the nucleus $^{12}$C.
For a comparison, results obtained with dropping out
the attractive (i.e.\ real) part of the $\eta A$ potential
are also shown.\protect\cite{sokol98}}
\label{fig:spectral_function}
\end{figure}

Theoretical estimates given in Refs.~\cite{lvov98,sokol98} show that
the binding effects lead to a full dominance of the reaction mechanism
related with the multiple rescattering of $\eta$ and a formation of the
intermediate $\eta$-nucleus over a mechanism of nonresonance background
production of the $\pi N$ pairs in the subthreshold invariant-mass
region $\sqrt{s_{\pi N}} < m_{\eta} + m_N$ of the subprocess $\eta + N
\to \pi + N$.  A peak in the mass distribution of $\pi N$ is
theoretically expected in this region.  This is illustrated in
Fig.~\ref{fig:spectral_function} where the spectral function $S(E)$ of
the (kinetic) energy $E$ of $\eta$ in the nucleus is shown which is
proportional to the number of $\eta N$ collisions which the $\eta$
experiences when travels through the nucleus.  Another spectral
function, $S(E,q)$, shows a distribution of the produced $\pi N$ pairs
over their total energy $(E + m_{\eta} + m_N)$ and momentum $q$.  The
presence of the $\eta N$ attractive potential produces strong
enhancements in the spectral functions in the energy-momentum region
corresponding to the bound $\eta A$ states.

\begin{figure}[htb]
\centerline{
\psfig{file=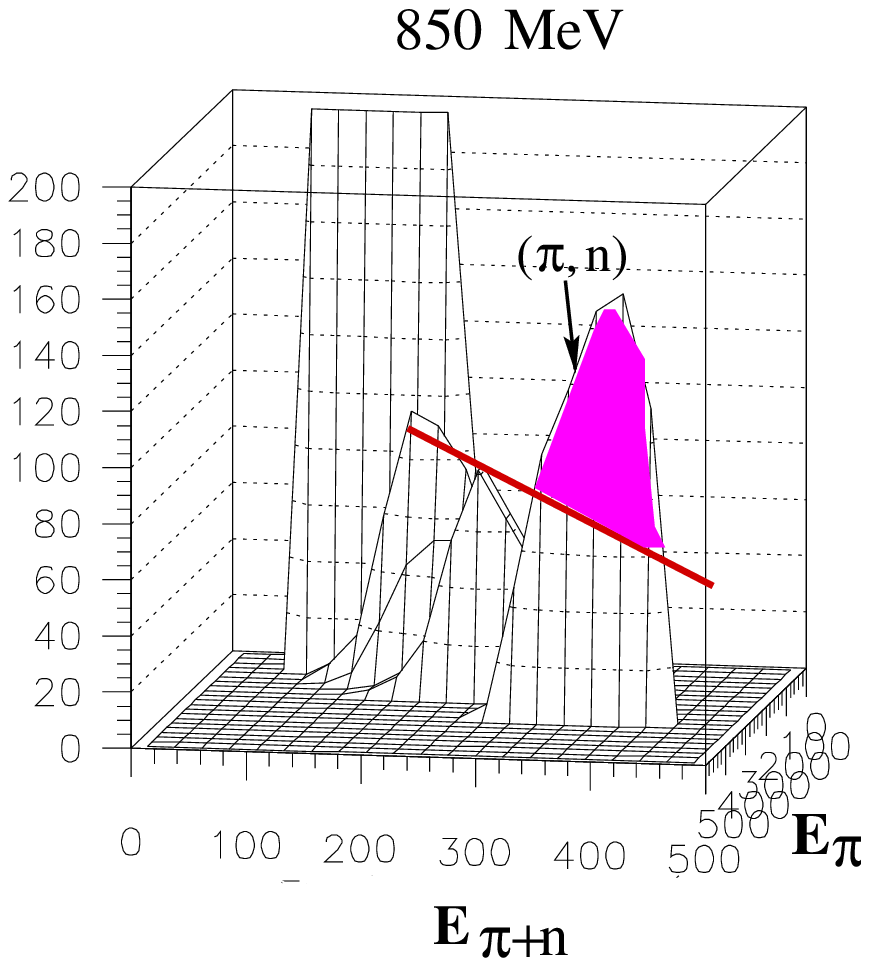,height=4.5cm} \qquad
\psfig{file=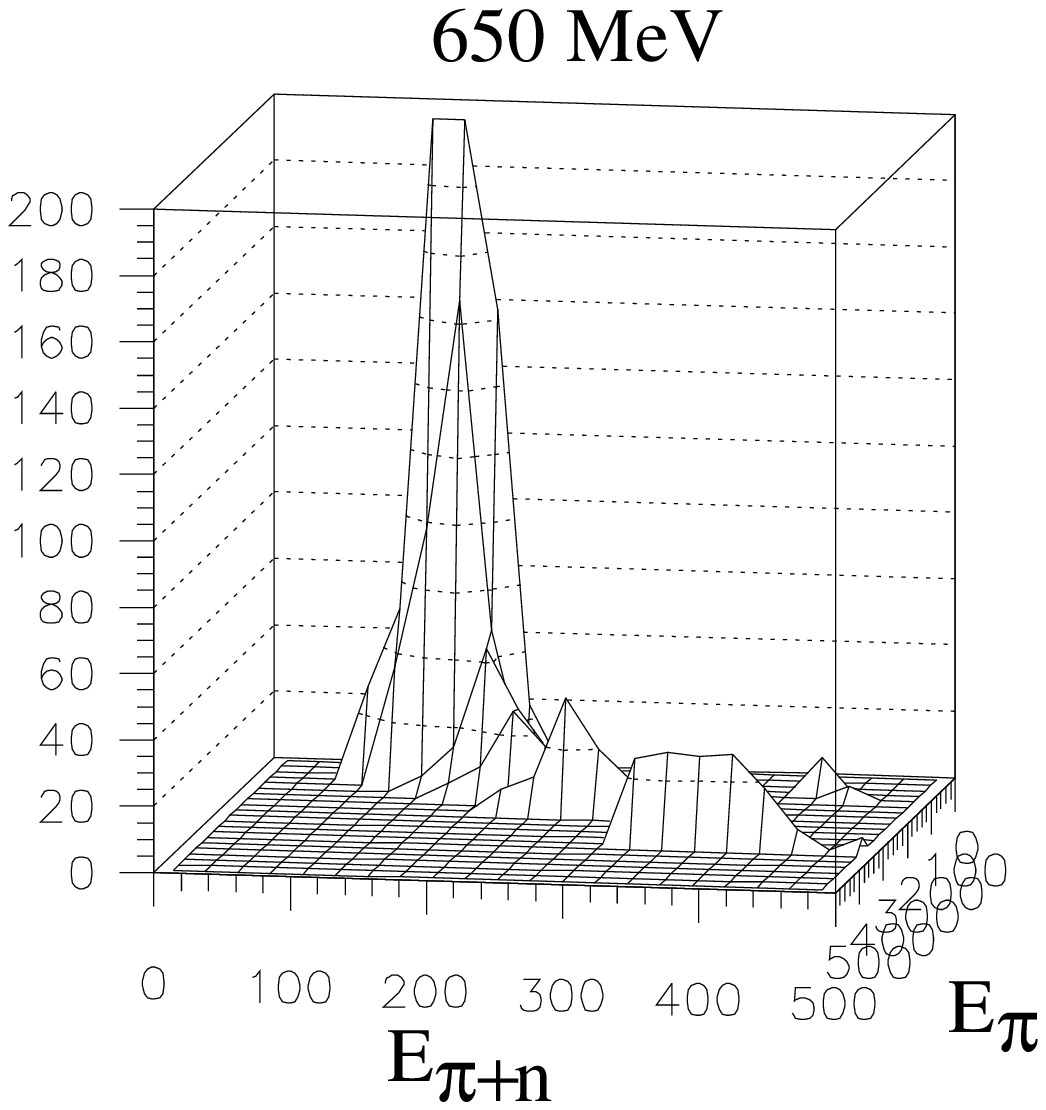,height=4.3cm}}
\caption{Distribution over the total kinetic energy of the $\pi^+n$ pairs
for the ``effect $+$ background" run
($E_{\gamma\,\rm max}=850$ MeV, the left panel) and for the
"background" run ($E_{\gamma\,\rm max}=650$ MeV, the right panel)
obtained after unfolding the raw spectra.}
\label{fig:Etot-2dim}
\end{figure}

The first experimental evidence for formation of the $\eta$-nuclei in
photoreactions was recently obtained in Ref.~\cite{sokol98}.  The
experiment was performed at the bremsstrahlung photon beam of the 1 GeV
electron synchrotron of Lebedev Physical Institute.  The end-point
photon energies $E_{\gamma max} = 650$ and 850 MeV were used which are
below and above $\eta$-photoproduction threshold off the free nucleon.
The reaction studied was
$$
   \gamma + {}^{12}{\rm C} \to p(n) + {}^{11}_{\eta}{\rm B}
   ({}^{11}_{\eta}{\rm C}) \to \pi^+ + n + X,
$$
in which energy-momentum correlations of $\pi^+$ and $n$ were studied.
Two time-of-flight scintillation spectrometers for detection of $\pi^+$
and $n$ were placed in opposite directions at $90^\circ$ with respect
to the photon beam.  Note that the $\pi^+ n$ pairs flying transversely
to the photon beam cannot be produced via the one-step reaction $\gamma
p \to \pi^+ n$ in the nucleus, whereas they naturally appear through an
intermediate production and annihilation of a slow $\eta$-meson.  Such
transversely-flying pairs have indeed been observed in the experiment.
Moreover, a resonance peak in the total energy of the $\pi^+ n$ pairs
has been found which appeared when the photon energy $E_{\gamma max} =
850$ MeV exceeded the $\eta$-production threshold (this is the
kinematics of ``effect$+$background"\cite{sokol98}) and did not appear
when $E_{\gamma max} = 650$ MeV (the kinematics of
``background"\cite{sokol98}).  See Fig.~\ref{fig:Etot-2dim}.  This peak
was interpreted as a manifestation of decays of the bound $\eta$'s in
the nucleus, i.e. a formation and decay of the $\eta$-mesic nuclei.

\begin{figure}[htb]
\centerline{\psfig{file=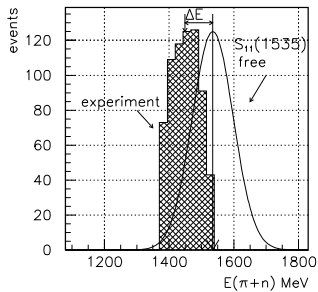,width=0.35\textwidth}}
\caption{Distribution over the total energy of the $\pi^+n$ pairs
after a subtraction of the background. }
\label{fig:Etot-1dim}
\end{figure}

Subtracting a smooth background and unfolding the measured (raw)
spectra with experimental resolutions, the 1-dimensional energy
distribution over the total energy of the $\pi^+ n$ pairs has been
found (Fig.~\ref{fig:Etot-1dim}). It develops a peak by $\sim90$ MeV
lying below the mass of the $S_{11}(1535)$ resonance and even below the
threshold energy $m_\eta + m_N=1486$ MeV, thus indicating a presence of
binding effects.  The width of the peak is about 150 MeV which is
compatible with the experimental resolution and broadening effects due
to the Fermi motion of nucleons and the bound $\eta$ in the
intermediate $\eta$-nucleus.

In prospect, studies of the $\eta$-mesic nuclei lying at the 
intersection of the nuclear physics and the physics of hadrons promise 
to bring a new information important for both the fields.  The 
$\eta$-mesic nuclei provide a unique possibility to learn interactions 
of $\eta$-mesons with nucleons and nucleon resonances, both free and in 
the nuclear matter.  Detecting and measuring the energy of the nucleon 
knocked out in the process of quasi-free $\eta$-production on nucleons 
in the nucleus, one can tag the energy of $\eta$ staying in the 
nucleus.  Used with the tagged photons technique, this opens a 
possibility to study an energy dependence of interactions between 
$\eta$ and nuclear constituents.

We hope to fulfil further photoproduction experiments in order to 
measure binding energies of $\eta$ in different nuclei in the 
$A=3{-}16$ mass range.  Since the $\eta$ energy levels and widths in 
nuclei depend on many important characteristics like, e.g., the $\eta 
N$ potential and the self-energy of the $S_{11}(1535)$ in the nuclear 
matter, the expected data might be very useful for further progress in 
understanding exotic nuclear systems and the behaviour of hadrons in 
nuclei.

\section*{Acknowledgment}
This work was supported by the Russian Foundation for Basic Research,
grant 99-02-18224.

\end{document}